\documentclass[aps,prl,preprint,superscriptaddress,amsmath,amssymb,]{revtex4-1}
\newcommand{\degre}{^\circ}
\usepackage{graphicx}% Include figure files
\begin{document}

%\preprint{}

%\title{Near-field energy transfer by magneto-electric interference}
%\title{Photon harvesting by evanescent waves outgoing from very narrow grooves}
%\title{Role of the magneto-electric interference in the light funneling mechanism}

\title{Giant field enhancement by funneling effect into sub-wavelength slit-box 
resonators} 
%\title{Magneto-electric interference as the source of light funneling}
%\title{Light funneling mechanism}



\author{Paul Chevalier}
%\email{paul.chevalier@onera.fr}
%\thanks{}
\affiliation{Laboratoire de Photonique et de Nanostructures (LPN-CNRS),
Route de Nozay, 91460 Marcoussis, France}
\affiliation{ONERA, The French Aerospace Lab, 91761 Palaiseau, France}

\author{Patrick Bouchon}
\email {patrick.bouchon@onera.fr}
%\thanks{}
\affiliation{ONERA, The French Aerospace Lab, 91761 Palaiseau, France}

\author{Riad Ha{\"i}dar}
%\email{riad.haidar@onera.fr}
%\thanks{}
\affiliation{ONERA, The French Aerospace Lab, 91761 Palaiseau, France}
\affiliation{\'Ecole Polytechnique, D\'epartement de Physique, 91128 Palaiseau, France}

\author{Fabrice Pardo}
%\email{fabrice.pardo@lpn.cnrs.fr}
%\thanks{}
\affiliation{Laboratoire de Photonique et de Nanostructures (LPN-CNRS),
Route de Nozay, 91460 Marcoussis, France}


\date{\today}

\begin{abstract}
Inspired by the acoustic Helmholtz resonator, we propose a slit-box
electromagnetic nanoantenna able to concentrate the energy of an
incident beam into surfaces a thousand times smaller than with a
classical lens.
This design gives birth to giant field intensity enhancement in hot
volume, throughout the slit. It reaches $10^4$ in the
visible up to $10^8$ in the THz range even with focused beams thanks
to an omnidirectional reception.
These properties could target applications requiring extreme light
concentration, such as SEIRA, non-linear optics and biophotonics. 
\end{abstract}

\pacs{}
%\keywords{}

\maketitle

Optical nanoantennas are of great importance for harvesting light from the free space to deep-subwavelength volume and to reach a strong electric field enhancement~\cite{lerosey2007focusing,grbic2008near,zhang2008plasmonic,novotny2011antennas,biagioni2012nanoantennas,barnes2003surface,greffet2005nanoantennas,schuller2010plasmonics}.
They have a wide range of applications including surface enhanced Raman scattering (SERS)~\cite{doi:10.1021/nl403030g}, photovoltaics and photodetection~\cite{knight2011photodetection}, non-linear optics~\cite{kauranen2012nonlinear}, bio-sensing~\cite{liu2011nanoantenna}, surface enhancement infrared absorption (SEIRA)~\cite{dregely2013vibrational,bochterle2012angstrom,adato2009ultra}, and thermal emission~\cite{schuller2009optical}. Field intensity enhancement up to $10^3$ was 
observed using subwavelength particles~\cite{muhlschlegel2005resonant} and an enhancement up to $10^6$ is expected inside self-similar chains~\cite{li2003self}. 
Inspired by radiofrequency designs, various nanoantennas have been engineered to improve the field enhancement in larger volumes, such as bow-tie~\cite{schuck2005improving,kim2008high}, 
quarter-wave or Yagi-Uda~\cite{PhRvL-Novotny2007ews,kosako2010directional,dregely20113d} antennas. However, the antennas previously proposed either concentrate the electric field only in {\em{hot spots}}, or are plagued by their directivity which prevents them to be used with focused beam. The key issue to further enhance the electromagnetic field is thus to design an antenna having a large angular acceptance as well as a wide cross-section. 
Here, we present a slit-box resonator inspired by the acoustic Helmholtz resonator~\cite{von1896theorie}, which is able to omnidirectionally funnel all the light through its aperture regardless of its width, leading to million fold electric field intensity enhancement in a {\em{hot volume}}.
This design, allows to tune independently the slit (width: $w$, height: $h$) and the cavity geometries for a given resonance wavelength $\lambda$ and we demonstrate that it leads to a giant enhancement factor driven by the ratio $\lambda^2/(hw)$. This enhancement factor only depends slightly on the angle of incidence which permits to use this antenna at the focal spot of a classical lens to further strongly enhance the electric field intensity.
This giant field intensity enhancement opens great perspectives of application for devices based on SEIRA or nonlinear optics which used focused beam.

The funneling by magneto-electric interference was shown to efficiently collect light into
2D sub-wavelength slits with a near-$\lambda$ cross section and a large angular acceptance~\cite{pardo2011light,bouchon2011total,PhysRevLett.107.163902,chevalier2012funneling}.
In the following, the funneling effect is applied to a single Helmholtz-like nanoantenna at the focal point of a focused beam. 
Our electromagnetic structure is made of a metallic material and is 
long enough along the 
\textbf{y} direction to be considered as infinite. It consists in a box (width 
$w_b$, height $h_b$) and a slit (width $w_s$, height $h_s$) as shown in 
Fig.~1(a). 
A beam is focused on one of these structures where almost all the incident 
energy 
is funneled.
Exact Maxwell computations \cite{bouchon2010bmm} of this 
slit-box resonator are performed by considering that the metallic walls are 
made 
of gold.
Its dielectric function is computed from the 
Drude model $\varepsilon_{\mathrm{Au}}(\lambda) =1 - 
\left[\left(\lambda_p/\lambda + i 
\gamma\right)\lambda_p/\lambda\right]^{-1}$ which matches well the experimental 
data in the infrared domain for $\lambda_p = 159 \, \mathrm{nm}$ and $\gamma = 
0.0077$ \cite{palik1985handbook}. In the following we consider a box of width 
$w_{b} = 
49 \, \mathrm{nm}$ and of height $h_{b} = 49 \, \mathrm{nm}$,
and a slit of width $w_{s} = 2.5 \, \mathrm{nm}$ and of height $h_{s} = 11
\, \mathrm{nm}$ \cite{quantum} that is firstly studied at $\lambda=1.7 \,  \mu 
\mathrm{m}$.

\begin{figure*}[htp]
\centering
  \includegraphics[width=0.9\textwidth]{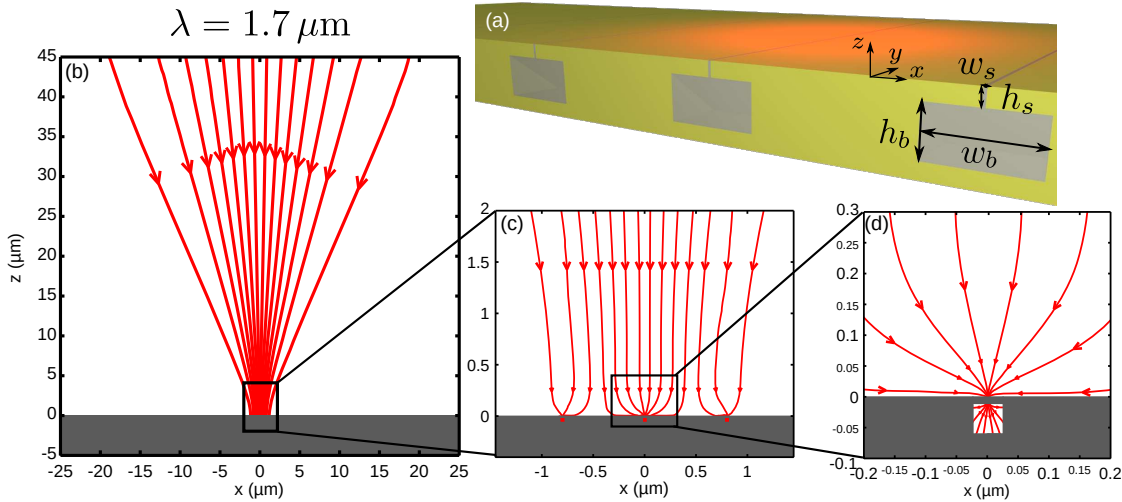}
  \caption{\label{fig:fig1}
(a) A beam is focused on a structure made of a slit of width $w_s=2.5 \, 
\mathrm{nm}$ and height $h_s=11 \, \mathrm{nm}$ ended by a box of width $w_b=49 
\, \mathrm{nm}$ and height $h_b=49 \, \mathrm{nm}$.
(b) Streamlines of the beam focused into the structure, the beam waist is about 
$\lambda= 1.7 \, 
\mu \mathrm{m}$ wide.
(c) The beam is focused on one individual structure that collects about 60 \% 
of the incoming energy. The remaining energy is fully collected by the 
neighboring slits. 
(d) Streamlines of the Poynting vector in the vicinity of the structure, showing 
that the magneto-electric interference redirects the incident energy inside the 
resonator~\cite{pardo2011light}.
}
\end{figure*}

The behavior of the structure has been simulated by focusing a beam onto a 
slit-box structure. Technically the structure is periodized with a period $d=0.8 
\, \mu \mathrm{m}$ slightly smaller than the wavelength (see supplementary 
materials). The streamlines of the Poynting vector plotted at a large scale in 
Fig.~\ref{fig:fig1}(b) show that the incident energy is focused into a beam limited 
by diffraction (see Fig.~\ref{fig:fig1}(c)) where 90\% of the energy is concentrated in a 
$\lambda$-wide spot
(here the wavelength is $\lambda=1.7 \, \mu \mathrm{m}$). Next, in the 
near field ($z<500 \, \mathrm{nm}$, see Fig.~\ref{fig:fig1}(d)), all the focused energy 
is funneled towards the opening of the resonators where it is absorbed on the 
metallic 
sidewalls. In fact, given the size of the focal spot, 60\% of the total energy 
is absorbed by the central resonator while the remaining energy is equally 
absorbed by the two surrounding resonators. Since the incoming energy goes 
through the slit of the resonator, the electric field intensity is strongly 
enhanced in the whole slit volume.

\begin{figure*}[htp]
\centering
  \includegraphics[width=0.8\textwidth]{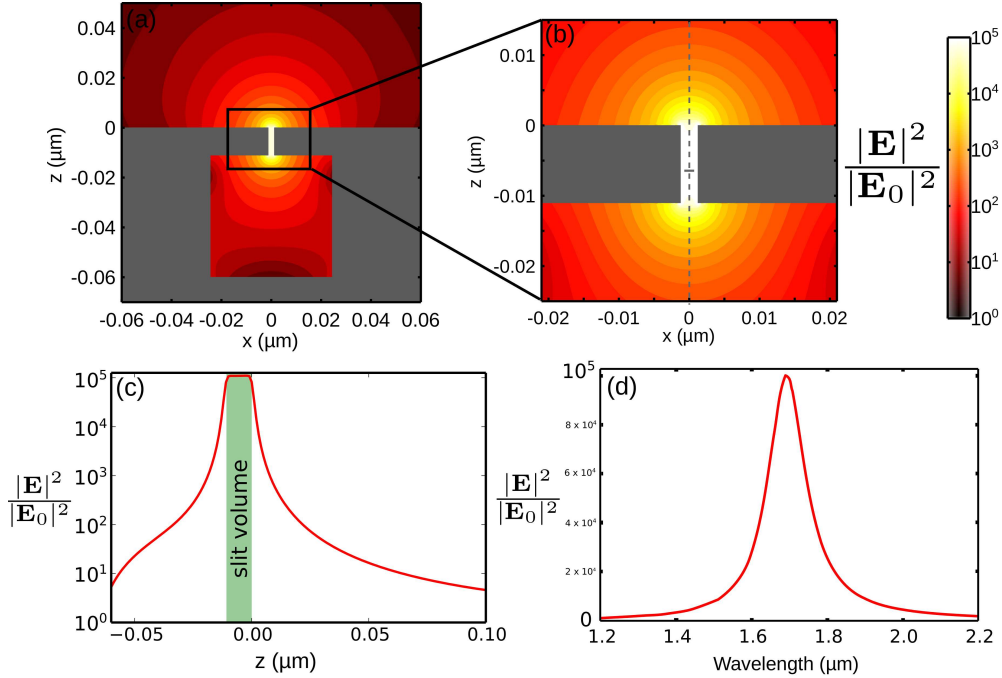}
  \caption{\label{fig:fig2}
 The electric field intensity enhancement is shown at different scales for an 
incident focused beam.
(a) Global field intensity enhancement at $\lambda=1.7\, \mu \mathrm{m}$.
(b) Close view in the vicinity of the slit. 
(c) Field intensity enhancement along the $z$-axis, at the center of the 
structure ($x=0$). A rather homogeneous enhancement of $10^5$ is found in the 
whole slit volume.
(d) The field intensity enhancement in the center of the slit is plotted at 
different 
wavelengths. We see that $\lambda=1.7\, \mu \mathrm{m}$ is a resonance 
wavelength that provide a maximal enhancement. The quality factor of this 
resonance is about 10.
}
\end{figure*}

The distribution of the electric field intensity ($|E|^2$) normalized by the 
incident 
field intensity ($|E_0|^2$) in the central structure is presented in 
Figs.~\ref{fig:fig2}(a)-\ref{fig:fig2}(b) 
at different zoom levels and shows that a field intensity enhancement of $10^5$ 
is reached with this structure. In Fig.~\ref{fig:fig2}(c), it is shown that this enhancement is confined to the slit 
volume: the field intensity enhancement stays constant along the height of the 
slit, but outside its volume the electric field intensity quickly drops by 
losing 4 orders of magnitude over about $50 \, \mathrm{nm}$.
Noteworthily, this profile of the electric field departs from the behaviour of a Fabry-Perot slit, and it adds to the fact that the slit-box resonator does not exhibit harmonics features (see Fig. S1 in supplemental materials.)%
The quality factor is $Q \simeq 10$ for a resonance wavelength of $\lambda_R = 
1.7\,  \mu \mathrm{m}$ as shown in Fig.~\ref{fig:fig2}(d).
In the supplementary materials, we show that the slit-box structure is fairly well described by a LC resonator model
(the capacitor being the slit, and the inductance being the box that acts as a magnetic energy accumulator), giving an analytic formula for
the resonance wavelength: $\lambda_R \simeq 2 \pi n_s \sqrt{w_b h_b h_s/w_s}$ 
where $n_s$ is the refractive index of the dielectric material filling the 
slit. It must be emphasized that, as expected from the inductive nature of the 
box, its dielectric filling plays no role on the resonance wavelength.

\begin{figure}[htp]
\centering
  \includegraphics[width=0.4\textwidth]{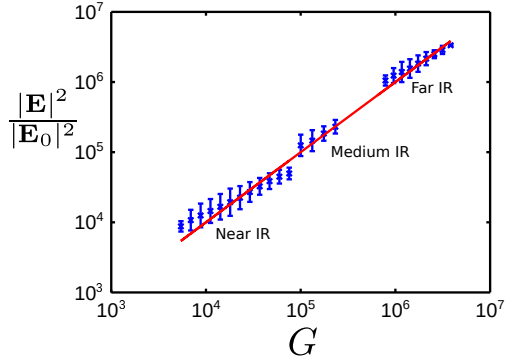}
  \caption{\label{fig:fig3} Simulations for different size of structure have 
been realized with $\varepsilon_s=1$. For each structure the field intensity 
enhancement for a plane incident wave is computed and compared to the model 
provided here. We see a strong correlation between the simulations results and 
the formula provided in the manuscript for the different domains of frequency. 
Three sets of simulations have been made, the first in the near IR domain 
($\lambda$ between $2 \, \mu \mathrm{m}$ and $8 \, \mu \mathrm{m}$), the second 
in the medium IR ($\lambda$ between $10 \, \mu \mathrm{m}$ and $25 \, \mu 
\mathrm{m}$)  and the last in the far IR domain ($\lambda$ between $30 \, \mu 
\mathrm{m}$ and $80 \, \mu \mathrm{m}$)}
\end{figure}

For an incident wave polarized with the electric field perpendicular to the 
slits, the maximum field intensity enhancement $G_{max}$ in the slit at the resonance can be expressed as:
\begin{equation}
G_{max}=\frac{|E_{max}|^2}{|E_0|^2} = \frac{Q}{2\pi} \frac{\lambda^2}{h_s w_s },
\end{equation}

where $Q$ is the quality factor of the resonance, $\lambda$ is the wavelength. This formula stems from the 
expression of the stored energy and the dissipated energy inside the resonators
(see supplementary 
materials). It stands for a plane wave normally incident onto a perfectly 
impedance-matched structure, exhibiting total absorption (i.e. neither 
scattered nor reflected fields).
In a more general situation, where the energy absorption efficiency $A$ is different from $1$, or when the period $d$ is different from $\lambda$, the field intensity enhancement writes $G=\frac{Q}{2\pi} \frac{A \lambda d}{h_s w_s}$.
In figure~\ref{fig:fig3} the enhancement factor has been calculated for 
variously shaped structures ($\varepsilon_s=1$) exhibiting resonances from the near infrared to the far infrared and is plotted as a function of $G$. This figure shows the good agreement between this analytic model and the values obtained through numerical simulations in the different domains of frequencies.

\begin{figure}[htp]
\centering
  \includegraphics[width=0.4\textwidth]{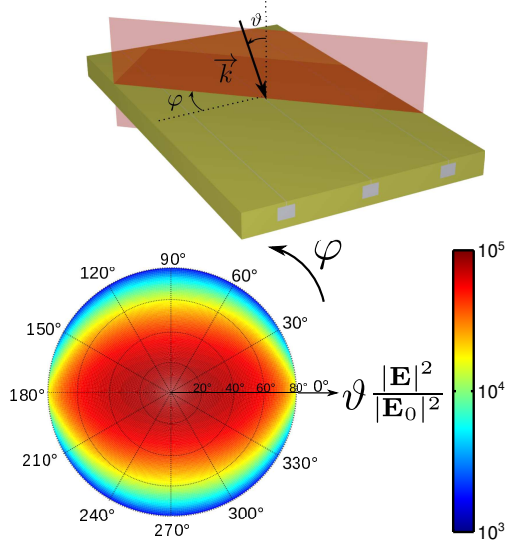}
  \caption{\label{fig:fig4} The electric field intensity enhancement is 
computed for an incident plane wave (polarized parallel to the slits) with 
different angles of incidence and different planes of incidence. The structure 
is the same as described in the manuscript. (a) The plane of incidence is rotated 
by an angle $\varphi$, the angle of incidence is $\vartheta$. (b) The results of 
the field intensity enhancement is represented with polar coordinates on the 
right. The field intensity enhancement remains above $5\times 10^4$ for angles 
of incidence below $40 \degre$, it is above $10^4$ for angle of incidence below 
$65\degre$.}
\end{figure}

The angular dependence of the enhancement is studied in Fig. \ref{fig:fig4}. The general configuration of conical incidence is described in Fig. \ref{fig:fig4}(a), the light is impinging on the antenna with an angle $\vartheta$ in the plane defined by the azimuthal angle $\varphi$ and is polarized so that the magnetic field is in the $y-z$ plane. The enhancement G as a function of both angles is plotted in Fig. \ref{fig:fig4}(b). It remains above $5.10^4$ for angle of incidence below $35^\circ$ which demonstrates the omnidirectional behaviour of the resonator.  
Due to its high angular tolerance,
the enhancement G would only be slightly reduced when the structure is placed at the focal spot
of a lens. In this case, the electric field intensity $|E|^2$ would be similar to the one 
obtained if, beyond the limits of diffractive optics, a beam of given power was focused in 
a spot of area $\lambda^2/G$ instead of $\lambda^2$. Noteworthily, this intensity is uniformly 
obtained  in a ``hot volume'' of height $h_s$ and area $\lambda w_s \gg \lambda^2/G$.

Very strong field intensity enhancements can be reached not only in the infrared range, but in the whole spectrum from visible to THz domain.
At higher wavelengths, losses in the metal are weaker than in the near 
IR, so the quality factor of the resonance is greater. Therefore,
higher field intensity enhancements will be reached at longer wavelengths.
As shown in the supplementary materials, such structures allow higher 
enhancement factors for the intensity of 
the electric field than in the literature: 
$10^5$ vs. 
$10^3$~\cite{muhlschlegel2005resonant,dregely2013vibrational,adato2009ultra,
schuller2010plasmonics} in the 
near IR or visible, $10^6$ vs. $10^4$ in the far IR~\cite{feuillet2013strong}, 
and $10^8$ vs. $10^6$ in the THz domain~\cite{seo2009terahertz}.

%%conclusion
The slit-box structure allows to funnel nearly all the incident energy through the 
arbitrarily narrow aperture of the slit leading to giant enhancement of the 
electric field in the whole slit volume. These appealing properties are very 
promising for light matter interactions such as in photodetection, SERS, SEIRA 
and non-linear optics.


\begin{acknowledgments}
We are thankful to J.-J. Greffet for precious discussions concerning this manuscript.
We acknowledge financial support from the ONERA through the SONS project.

\end{acknowledgments}

\end{document}